\documentclass[12pt,a4paper,titlepage]{article}
\usepackage{amsmath,color}

\def\slash#1{\not\!\!#1}
\renewcommand{\eqref}[1]{(\ref{#1})}

\usepackage{amsmath,amssymb}

\usepackage[utf8]{inputenc}
\usepackage[top=50pt,bottom=50pt,left=68pt,right=66pt]{geometry}

\definecolor{hyprf}{cmyk}{1,0.5,0,0}

\def\VEV#1{\left\langle #1\right\rangle}        

\begin{document}

\thispagestyle{empty}

{\hbox to\hsize{
\vbox{\noindent September 2018 \hfill IPMU18-0131 }}
\noindent  ~revised version \hfill WU-HEP-18-09}

\noindent
\vskip2.0cm
\begin{center}

{\Large\bf Massive vector multiplet with Dirac-Born-Infeld 
\vglue.1in and new Fayet-Iliopoulos terms in supergravity 
}

\vglue.3in

Hiroyuki Abe~${}^{a}$, Yermek Aldabergenov~${}^{b,c}$, Shuntaro Aoki~${}^{a}$, and  Sergei V. Ketov~${}^{b,c,d}$
\vglue.1in

${}^a$~Department of Physics, Waseda University, Tokyo 169-8555, Japan\\
${}^b$~Department of Physics, Tokyo Metropolitan University, \\
Minami-ohsawa 1-1, Hachioji-shi, Tokyo 192-0397, Japan \\
${}^c$~Research School of High-Energy Physics, Tomsk Polytechnic University,\\
2a Lenin Ave., Tomsk 634050, Russian Federation \\
${}^d$~Kavli Institute for the Physics and Mathematics of the Universe (IPMU),
\\The University of Tokyo, Chiba 277-8568, Japan \\
\vglue.1in

abe@waseda.jp, shun-soccer@akane.waseda.jp, aldabergenov-yermek@ed.tmu.ac.jp, ketov@tmu.ac.jp
\end{center}

\vglue.3in

\begin{center}
{\Large\bf Abstract}
\end{center}
\vglue.1in
\noindent  We propose a four-dimensional  $N=1$ supergravity-based Starobinsky-type inflationary model in terms of a single massive vector multiplet, whose action includes the Dirac-Born-Infeld-type kinetic terms and a generalized (new) Fayet-Iliopoulos-type term, without gauging the R-symmetry. The bosonic action and the scalar potential are computed. 
The inflaton is the superpartner of the Goldstino in our model, and supersymmetry is spontaneously broken after inflation by the D-type mechanism, whose scale is related to the value of the cosmological constant.

\newpage


\section{Introduction}

Supergravity is the natural framework for unification of bosons and fermions, and for unification of elementary particles with gravity. On the one hand, it is possible (though non-trivial) to unify the dark matter (as the lightest supersymmetric particle), the dark energy (as the positive cosmological constant) and the cosmological inflation (with the inflaton scalar field having the proper scalar potential) in supergravity. On the other hand, supergravity emerges as the low-energy effective theory from (compactified) superstrings (quantum gravity), and can be connected
to the Standard Model at the electro-weak scale. All phenomenological applications of supergravity require (spontaneous)
supersymmetry breaking and a non-vanishing gravitino mass.

In supersymmetry (SUSY), the inflaton should belong to a supermultiplet. A spontaneous SUSY breaking implies the existence of the spin-1/2 Goldstino that should also belong to a supermultiplet. In the literature of the inflationary model building based on supergravity, one usually assumes that the inflaton belongs to a chiral multiplet and the Goldstino belongs to another chiral multiplet \cite{Kawasaki:2000yn,Kallosh:2010ug,Kallosh:2010xz}, whose K\"ahler potential and superpotential can be appropriately chosen "by hand" \cite{Yamaguchi:2011kg}.~\footnote{We assume the existence of 
a stable vacuum after inflation, and ignore run-away solutions \cite{Gasperini:2001pc}.}  
 This gives rise to {\it four} real physical scalars and the need to distinguish the inflaton among them, while stabilizing the remaining three scalars during  a single-field inflation. It can be done in many ways, thus reducing the predictive power.

This freedom of choice can be reduced by minimizing the number of the physical degrees of freedom involved. The 
inflaton and Goldstino chiral multiplets can be identified, which leads to the viable and more economic inflationary models based on supergravity \cite{Ketov:2014qha,Ketov:2014hya,Ketov:2015tpa,Schmitz:2016kyr}.
 
It is also possible to employ a massive {\it vector} multiplet \cite{VanProeyen:1979ks} that has only {\it one} physical scalar to play the role of the inflaton, and then to identify its fermionic superpartner with the Goldstino, as the truly minimal option. This opportunity was investigated in \cite{Farakos:2013cqa,Ferrara:2013rsa}, where it was found that it is flexible enough to accommodate a cosmological inflation with any values of the Cosmic Microwave Background (CMB) radiation tilts $n_s$ and $r$. However, it was also observed that SUSY is necessarily restored after inflation in this class of the supergravity-based  inflationary models, which requires an extra mechanism of spontaneous SUSY breaking after inflation for reheating and viable phenomenology of particles. A solution to this problem was proposed in \cite{Aldabergenov:2016dcu,Aldabergenov:2017bjt}, by adding a chiral (Polonyi) multiplet with
a linear superpotential.

It is, therefore, the good question: is it possible to get rid of Polonyi multiplet, but still describe a viable cosmological inflation together with a spontaneous SUSY breaking after inflation, by using only a {\it single} (massive) vector multiplet? The affirmative answer apparently requires extra tools in supergravity theory, beyond the standard ones.

In this paper we employ the new supergravity construction that includes the following theoretical resources (tools):
\begin{itemize}
\item the manifest ({\it linearly} realized) local $N=1$ supersymmetry,
\item the inflaton and the Goldstino in a single (massive) $N=1$ vector multiplet,
\item the kinetic terms of the vector multiplet have the Dirac-Born-Infeld (DBI) structure  inspired by superstrings and D-branes \cite{Born:1934gh,Fradkin:1985qd,Leigh:1989jq,Ketov:2001dq},
\item the new Fayet-Iliopoulos (FI) terms in supergravity, that do not require gauging the R-symmetry
\cite{Cribiori:2017laj,Kuzenko:2018jlz,Antoniadis:2018cpq,Farakos:2018sgq,Aldabergenov:2018nzd},
\item a constant superpotential.
\end{itemize}

The manifest SUSY has the advantage of a straightforward addition of quantum corrections. The Goldstino as the superpartner of the inflaton is the minimal option. The DBI structure introduces the new (BI) scale into our model, that is arguably between the Grand Unification (GUT) scale and Planck scale. 

The use of a constant (field independent) FI term \cite{Fayet:1974jb} is highly restrictive in supergravity, because its  (old) standard construction (via Noether procedure)  required gauging of the R-symmetry \cite{Freedman:1976uk,Binetruy:2004hh}. However, when assuming a nonvanishing  vacuum expectation value (VEV) of the auxiliary $D$-component of the vector multiplet from the very beginning, one can introduce other (new) FI terms 
\cite{Cribiori:2017laj,Kuzenko:2018jlz,Antoniadis:2018cpq,Farakos:2018sgq,Aldabergenov:2018nzd} that do not require gauging the R-symmetry.

We consider only the Starobinsky-like inflationary models for definiteness and because they are most natural in
our construction. As regards Starobinsky inflation and its realizations in supergravity, see e.g., 
\cite{Starobinsky:1980te,Ketov:2012jt,Ketov:2013dfa,AIKK}.

Our paper is organized as follows. Our technical setup, based on the superconformal tensor calculus, is briefly reviewed in Sec.~2. In Sec.~3 we propose the new supergravity action, and compute its bosonic part that includes the scalar potential. In Sec.~4 we apply our construction to the Starobinsky-like inflation and spontaneous SUSY breaking. Our conclusion is Sec.~5. In Appendix A we describe our supergravity actions in terms of the superfields defined in curved superspace. In Appendix B we briefly study the impact of a constant superpotential.

\section{Our setup}

In the main body of our paper (except of Appendix A) we use the conformal $N=1$ supergravity techniques~\cite{Kaku:1978nz,Kaku:1978ea,Townsend:1979ki, Kugo:1982cu,Kugo:1983mv}, and follow the notation and conventions of Ref.~\cite{Freedman:2012zz}. In addition to the symmetries of Poincar\'e supergravity, one also has the gauge invariance under dilatations, conformal boosts and
$S$-supersymmetry, as well as under $U(1)_A$ rotations. The gauge fields of dilatations and  
$U(1)_A$ rotations are denoted by $b_{\mu}$ and $A_{\mu}$, respectively. A multiplet of conformal supergravity has charges with respect to dilatations and $U(1)_A$ rotations, called Weyl and chiral weights, respectively, which are denoted by pairs $({\rm{Weyl\ weight, chiral\ weight}})$ in what follows. 

A chiral multiplet has field components 
\begin{align}
S=\{S,P_L\chi, F\}, \label{chiral}
\end{align}
where $S$ and $F$ are complex scalars, and $P_L\chi$ is a left-handed Weyl fermion ($P_L$ is the chiral projection operator). As regards a general multiplet, it has  
\begin{align}
\Phi=\{ \mathcal{C}, \mathcal{Z}, \mathcal{H}, \mathcal{K}, \mathcal{B}_{a}, \Lambda , \mathcal{D}\} , \label{general}
\end{align}
where $ \mathcal{Z}$ and $\Lambda $ are fermions, and the other fields are complex scalars.

The (gauge) field strength multiplet $W$ has the weights $(3/2,3/2)$ and the following components: 
\begin{align}
\bar{\eta } W=\left\{ \bar{\eta }P_L\lambda ,\frac{1}{\sqrt{2}}\left(  -\frac{1}{2}\gamma _{ab}F^{ab}+iD\right) P_L\eta  ,\bar{\eta  } P_L\slash{D}\lambda \right\} ,
\end{align}
where $\eta $ is the dummy spinor. $F_{ab}=\partial_aB_b-\partial_bB_a$ is the Abelian field strength,  and
$\lambda $ and $D$ are Majorana fermion and the real scalar, respectively. The related expressions of the multiplets $W^2$ and $W^2\bar{W}^2$, which are embedded into the chiral multiplet~$\eqref{chiral}$ and the general multiplet~$\eqref{general}$, respectively, are 
\begin{align}
&W^2=\left\{ \cdots , \cdots, \cdots+\frac{1}{2} (FF-F\tilde{F})-D^2\right\} ,\\
&W^2\bar{W}^2=\left\{ \cdots , \cdots, \cdots,\cdots , \cdots, \cdots, \cdots+\frac{1}{2}| (FF-F\tilde{F})-2D^2|^2\right\} ,
\end{align}
where we have omitted the fermionic terms (denoted by dots) for simplicity. 

In addition, we use the book-keeping notation  $FF=F_{ab}F^{ab}$ and $\tilde{F}^{ab}\equiv -\frac{i}{2}\varepsilon ^{abcd}F_{cd}$.

We also need another chiral multiplet
\begin{align}
\Sigma \left( \bar{W}^2/|S_0|^4\right) =\left\{ -\frac{ (\frac{1}{2}FF+\frac{1}{2}F\tilde{F}-D^2)}{|S_0|^4} +\cdots , \cdots, \frac{F_0}{|S_0|^4S_0}(FF+F\tilde{F}-2D^2)+\cdots\right\} , \label{D2W2}
\end{align}
where $\Sigma$ is the chiral projection operator \cite{Kugo:1982cu, Kugo:1983mv}. The argument of 
$\Sigma$ requires the specific Weyl and chiral weights: in order for $\Sigma \Phi$ to make sense, $\Phi$ must satisfy $w-n=2$, where $(w,n)$ are Weyl and chiral weights of $\Phi$. We get the correct weights by inserting the factor $|S_0|^4$, where $S_0$ is the chiral compensator of weights $(1,1)$.\footnote{The conformal supergravity compensators are distinguished from the physical matter supermultiplets in our notation by attaching the subscript $0$ to the former.} Equation~$\eqref{D2W2}$ is the conformal  supergravity counterpart of the superfield $\bar{D}^2\bar{W}^2$. 

The covariant derivative of $W$ is given by \cite{Kugo:1983mv}
\begin{align}
\mathcal{D}W=\{ -2D, \cdots , \cdots, \cdots,\cdots , \cdots,\cdots \} 
\end{align}
and has weights $(2,0)$. The dots in the higher components also include some bosonic terms, but we do not write  them here for simplicity (see Ref.~\cite{Cribiori:2017laj} for their explicit expressions).

A massive vector multiplet $V$ has field components 
\begin{align}
V=\{C,Z,H,K,B_a, \lambda ,D\}~,
\end{align}
while all of them are either real (bosonic) or Majorana (fermionic). The weights of $V$ are $(0,0)$.

The bosonic part of the F-term invariant action 
\begin{align}
[S]_F=\int d ^{4}x\sqrt{-g}\frac{1}{2}\left( F+\bar{F}\right), \label{Fformula}
\end{align}
 can be only applied when the $S$ has weights $(3,3)$. The bosonic part of the D-term of a real multiplet $\phi $ of weights $(2,0)$ reads
\begin{align}
 [\phi ]_D= \int d^{4}x\sqrt{-g}\left( D_{\phi}-\frac{1}{3}C_{\phi}R(\omega)\right) ,\label{Dformula}
\end{align}
where $R(w)$ is (superconformal) Ricci scalar in terms of spacetime metric and $b_{\mu}$~\cite{Freedman:2012zz}. The  $C_{\phi}$ and $D_{\phi}$ are the first and the last components of $\phi$, respectively.

\section{Our action}

Having defined the multiplets and the compensators in Sec.~2, we propose the following action: 
\begin{align}
S=S_{V}+S_{DBI}+S_{FI}~, \label{total action}
\end{align}
where we have defined
\begin{align}
&S_V=\biggl[ |S_0|^2\mathcal{H}(V)\biggr] _D, \\
&S_{DBI}=-\frac{1}{2}[W^2]_F+ \biggl[ \frac{\alpha (V)}{|S_0|^4} \frac{W^2\bar{W}^2}{1-2\alpha (V) \mathcal{A}+\sqrt{1-4\alpha (V)\mathcal{A}+4\alpha (V)^2\mathcal{B}^2} }\biggr] _D~~,\\
&S_{FI}=\biggl[ |S_0|^2\mathcal{I}(V)\frac{W^2\bar{W}^2}{(\mathcal{D}W)^2(\bar{\mathcal{D}}\bar{W})^2}\mathcal{D}W\biggr] _D ~~,
\end{align}
in terms of three arbitrary real functions $\mathcal{H}$, $\mathcal{I}$, and $\alpha $ of the vector multiplet $V$.  In addition, we have introduced 
\begin{align}
\mathcal{A}=\bar{\Sigma }\left( \frac{W^2}{|S_0|^4}\right) +{\rm{h.c.}}, \ \ \mathcal{B}=\bar{\Sigma }\left( \frac{W^2}{|S_0|^4}\right) -{\rm{h.c.}},
\end{align} 
which have weights $(0,0)$. 

The supergravity theory (\ref{total action}) without the $S_{FI}$ term was
proposed and studied in Ref.~\cite{Abe:2015fha}. The new FI term above (see also \cite{Kuzenko:2018jlz,Aldabergenov:2018nzd}) represents its non-trivial extension. Our FI term is different from the one introduced in \cite{Cribiori:2017laj} 
because it has the different structure and includes arbitrary function (See Appendix A for details). In \cite{Antoniadis:2018cpq}, the FI term of \cite{Cribiori:2017laj} is applied to a D-term inflation, where the inflaton belongs to a (charged) chiral multiplet. Our FI term for a vector multiplet is also different from the other FI terms in terms of scalar multiplets \cite{Farakos:2018sgq}.

It is straightforward to calculate the bosonic terms of the action $\eqref{total action}$. They are given by~\footnote{The Lagrangian density is defined by $S=\int d^4 x \sqrt{-g} \;\mathcal{L}$.}
\begin{align}
\nonumber \mathcal{L}_V=&-\frac{1}{3}|S_0|^2\mathcal{H}R(\omega)+2\mathcal{H}\left( |F_0|^2-|D_aS_0|^2\right) +|S_0|^2\mathcal{H}_CD\\
\nonumber &+\frac{1}{2}|S_0|^2\mathcal{H}_{CC}\left( |N|^2-B_a^2-(D_aC)^2\right) \\
&+\biggl\{-\mathcal{H}_CNS_0\bar{F}_0+\mathcal{H}_Ci\left( B_a+iD_aC\right) S_0D^a\bar{S}_0+{\rm{h.c.}} \biggr\} ~, \label{LV}\\
\mathcal{L}_{DBI}=&\frac{|S_0|^4}{8\alpha}\biggl[ 1- \sqrt{1-\frac{8\alpha}{|S_0|^4}\left( D^2-\frac{1}{2}FF\right)+\frac{4\alpha^2}{|S_0|^8}(F\tilde{F})^2}\biggr]~, \label{LDBI}\\
\mathcal{L}_{FI}=&-\mathcal{I}|S_0|^2\frac{\left(  D^2-\frac{1}{2}FF\right)^2-\frac{1}{4}(F\tilde{F})^2}{4D^3}~, \label{LFI}
\end{align}
where $N\equiv H+iK$, and the subscript on $\mathcal{H}$ denotes the derivative with respect to $C$. The $D_a$ is the  superconformal covariant derivative~\cite{Freedman:2012zz}, 
\begin{align}
D_aS_0=\partial_a S_0-iA_aS_0-b_aS_0~, \quad D_aC=\partial_a C-2b_aC~.
\end{align}

To eliminate the extra symmetries of conformal supergravity against Poincar\'e supergravity, we impose the following superconformal gauge fixing conditions:
\begin{align}
&{\rm{D-gauge}} :\ -\frac{1}{3}|S_0|^2\mathcal{H}=\frac{1}{2}~~,\\
&{\rm{A-gauge}} :\ S_0=\bar{S}_0~~,\\
&{\rm{K-gauge}} :\ b_{\mu}=0~~,
\end{align}
which guarantee that the Ricci scalar in the supergravity action is canonically normalized.~\footnote{The gauge fixing condition of $S$-supersymmetry is irrelevant for bosonic terms.} Then the $R(\omega)$ becomes the usual Ricci scalar $R$. Under the above conditions, Eq.~$\eqref{LV}$ becomes
\begin{align}
\nonumber \mathcal{L}_V=& \frac{1}{2}R-\frac{3}{4\mathcal{H}^2}(\mathcal{H}_C^2-\mathcal{H}_{CC}\mathcal{H})(\partial_aC)^2  +\frac{3\mathcal{H}_{CC}}{4\mathcal{H}}B_a^2\\
\nonumber&+3A_a^2+\frac{3\mathcal{H}_C}{\mathcal{H}}A_aB^a+2\mathcal{H}|F_0|^2-\frac{3\mathcal{H}_{CC}}{4\mathcal{H}}|N|^2\\
&+\biggl\{ -\sqrt{\frac{-3}{2\mathcal{H}}}\mathcal{H}_C\bar{F}_0N+{\rm{h.c.}}\biggr\} -\frac{3\mathcal{H}_C}{2\mathcal{H}}D~.
\end{align} 

Integrating out the auxiliary fields $A_a$, $N$ and $F_0$ by using their (algebraic) equations of motion (except for the auxiliary field $D$) yieds~\footnote{The auxiliary fields $A_a$, $N$ and $F_0$ were not included in Eqs.~$\eqref{LDBI}$ and $\eqref{LFI}$, because they do not contribute to the bosonic action.}
\begin{align}
A_a=-\frac{\mathcal{H}_C}{2\mathcal{H}}B_a,\ \ N=F_0=0~. \label{eom_auxiliary}
\end{align}
Substituting them into Eq.~$\eqref{LV}$, we obtain
\begin{align}
\mathcal{L}_V=& \frac{1}{2}R-\frac{1}{2}\mathcal{J}_{CC}(\partial_aC)^2 -\frac{1}{2}\mathcal{J}_{CC}B_a^2+\mathcal{J}_CD~,
\end{align}
where $\mathcal{J}(C)\equiv -\frac{3}{2}\log \left( -\frac{2}{3}\mathcal{H}\right)$.

The full bosonic Lagrangian, before integration over the auxiliary field $D$, is thus given by
\begin{align}
\nonumber \mathcal{L}=& \frac{1}{2}R-\frac{1}{2}\mathcal{J}_{CC}(\partial_aC)^2 -\frac{1}{2}\mathcal{J}_{CC}B_a^2+\mathcal{J}_CD\\
\nonumber &+\frac{e^{4\mathcal{J}/3}}{8\alpha}\biggl[ 1- \sqrt{1-8\alpha e^{-4\mathcal{J}/3}\left( D^2-\frac{1}{2}FF\right)+4\alpha^2 e^{-8\mathcal{J}/3}(F\tilde{F})^2}\biggr]\\
&-\mathcal{I}e^{2\mathcal{J}/3}\frac{\left(  D^2-\frac{1}{2}FF\right)^2-\frac{1}{4}(F\tilde{F})^2}{4D^3}~~.
\end{align}

Let us consider the elimination of $D$ that is non-trivial. Its equation of motion is given by
\begin{align}
\nonumber &\mathcal{J}_C +\frac{1}{\sqrt{f^2-8\alpha e^{-4\mathcal{J}/3}D^2}}D\\
&-\mathcal{I}\frac{e^{2\mathcal{J}/3}}{4}\left(  1+\frac{FF}{D^2}-\frac{3}{4}\frac{(FF)^2-(F\tilde{F})^2}{D^4}\right) =0, \label{EOM D}
\end{align}
where we have introduced the function 
\begin{align}
f(F)\equiv \sqrt{1+4\alpha e^{-4\mathcal{J}/3}FF+4\alpha ^2e^{-8\mathcal{J}/3}(F\tilde{F})^2}~~.\end{align}
Hence, $D$ is a root of the 5th order polynomial, and it is impossible to solve (\ref{EOM D}) explicitly.

However, when the FI term vanishes, i.e., $\mathcal{I}=0$, Eq.~$\eqref{EOM D}$ takes the form
\begin{align}
\mathcal{J}_C +\frac{1}{\sqrt{f^2-8\alpha e^{-4\mathcal{J}/3}D^2}}D=0~, \label{EOM D0}
\end{align}
and its solution can be found as 
\begin{align}
D^{(0)}=\pm K f~, \label{sol 0}
\end{align}
where we have defined
\begin{align}
K(C)\equiv \sqrt{\frac{\mathcal{J}_C^2}{1+8\alpha \mathcal{J}_C^2 e^{-4\mathcal{J}/3}}}~~,
\end{align}
and $D^{(0)}$ stands for the solution at $\mathcal{I}=0$.~\footnote{Though the full theory is inconsistent in the
limit $\xi=0$ that also implies $\mathcal{I}=0$ for our choice of this function in (\ref{simplify}) below, Taylor expansion 
of the solution to (\ref{EOM D}) with respect to $\mathcal{I}$ and the $D^{(0)}$ are well defined. We always assume that
$\xi\neq 0$ and $\VEV{D}\neq 0$.} 

Since our interest is in what happens when $\mathcal{I}\neq 0$, we seek a perturbative solution to be connected to $D^{(0)}$ in the limit of $\mathcal{I}\rightarrow 0$.  We assume that the perturbative solution takes the form
\begin{align}
D=D^{(0)}+\mathcal{I}D^{(1)}+ \mathcal{O}(\mathcal{I}^2). \label{sol per}
\end{align}  
Substituting this into Eq.~$\eqref{EOM D}$ and considering the coefficient of $\mathcal{I}$, we obtain the equation
\begin{align}
\frac{4\mathcal{J}_C^3e^{-2\mathcal{J}/3}}{K^2}\frac{D^{(1)}}{D^{(0)}}+1+\frac{FF}{D^{(0)2}}-\frac{3}{4}\frac{(FF)^2-(F\tilde{F})^2}{D^{(0)4}}=0~, \label{EOM D1}
\end{align} 
where we have neglected the terms proportional to $\mathcal{I}^n, (n\geq 2) $, and have used the fact that $D^{(0)}$ satisfies Eq.~$\eqref{EOM D0}$. Note that the zero-th order equation with respect to $\mathcal{I}$ is trivially satisfied since $D^{(0)}$ is the solution when $\mathcal{I}=0$. From Eq.~$\eqref{EOM D1}$, we find 
\begin{align}
D^{(1)}=\mp \frac{e^{2\mathcal{J}/3}K^3}{4\mathcal{J}_C^3}f\left( 1+\frac{FF}{K^2f^2} -\frac{3}{4}\frac{(FF)^2-(F\tilde{F})^2}{K^4f^4} \right) .
\end{align} 

Hence, the bosonic Lagrangian up to the first order in $\mathcal{I}$ reads  
\begin{align}
\nonumber \mathcal{L}=& \frac{1}{2}R-\frac{1}{2}\mathcal{J}_{CC}(\partial_aC)^2 -\frac{1}{2}\mathcal{J}_{CC}B_a^2+\frac{e^{4\mathcal{J}/3}}{8\alpha } \left( 1\pm \frac{\mathcal{J}_C}{K} f\right) \\
&\mp \mathcal{I} \frac{e^{2\mathcal{J}/3}}{4}Kf \left( 1-\frac{FF}{K^2f^2} +\frac{(FF)^2-(F\tilde{F})^2}{4K^4f^4}\right)+\mathcal{O}(\mathcal{I}^2) . \label{onshell_action}
\end{align}
In particular, as regards the real scalar of the vector multiplet, $C$, we get its Lagrangian as
\begin{align}
&\mathcal{L}_C =-\frac{1}{2}\mathcal{J}_{CC}(\partial_aC)^2 -V,\\
&V= -\frac{e^{4\mathcal{J}/3}}{8\alpha } \left( 1\pm \frac{\mathcal{J}_C}{K} \right)  \pm \mathcal{I} \frac{e^{2\mathcal{J}/3}}{4}K  +\mathcal{O}(\mathcal{I}^2) .
\end{align} 

Fortunately, it is possible to compute the scalar potential $V(C)$ non-perturbatively, when
ignoring the $F$-terms of the vector field. Indeed,  when $F=0$, the $D$-equation $\eqref{EOM D}$ can be solved exactly, and its solution is given by
\begin{align} \label{Dsol}
D=\pm \sqrt{\frac{\left( \mathcal{J}_C-\frac{\mathcal{I}}{4}e^{2\mathcal{J}/3} \right)^2}{1+8\alpha e^{-4\mathcal{J}/3}\left( \mathcal{J}_C-\frac{\mathcal{I}}{4}e^{2\mathcal{J}/3} \right)^2 }}~~.
\end{align}
Therefore, the full scalar Lagrangian becomes 
\begin{align}
&\mathcal{L}_C =-\frac{1}{2}\mathcal{J}_{CC}(\partial_aC)^2 -V, \label{action_inflaton}\\
&V= -\frac{e^{4\mathcal{J}/3}}{8\alpha } \left( 1\pm \sqrt{1+8\alpha e^{-4\mathcal{J}/3}\left( \mathcal{J}_C-\frac{\mathcal{I}}{4}e^{2\mathcal{J}/3} \right)^2} \right) . \label{potential_inflaton}
\end{align} 
We choose the minus sign in Eq.~$\eqref{potential_inflaton}$ because it is the only option consistent with 
Eq.~(\ref{EOM D}).

Some comments are in order.

First, the perturbative solution $\eqref{onshell_action}$ allows us to investigate the sign in front of the vector kinetic terns $F^2$ in our action. In order to avoid ghosts, the sign should be negative,
\begin{align}
-\frac{1}{4K} \left( \mathcal{J}_C +\mathcal{I}e^{2\mathcal{J}/3} -2\alpha K^2\mathcal{I}e^{-2\mathcal{J}/3}\right) <0~, \label{cond_FF}
\end{align} 
which imposes the restriction on our functions.

Second, we can generalize our action (\ref{total action}) even further by adding a  constant superpotential $w$ as the additional term
\begin{align}
S_w=2[S_0^3w]_F~~,\label{constant_super}
\end{align} 
because there is no gauged $R$-symmetry in our approach. Then the extra bosonic part is 
\begin{align}
\mathcal{L}_w=3wS_0^2F_0+{\rm{h.c.}},
\end{align} 
and the superconformal gauge conditions lead to
\begin{align}
\mathcal{L}_w=-\frac{9}{2\mathcal{H}}F_0w+{\rm{h.c.}}
\end{align} 
Hence, the auxiliary fields equations of motion for $N$ and $F_0$ --- see Eq.~$\eqref{eom_auxiliary}$ --- change as 
\begin{align}
N=\sqrt{\frac{-8\mathcal{H}}{3}}\frac{\mathcal{H}_C}{\mathcal{H}_{CC}}F_0 \quad {\rm and} \quad
 F_0=\frac{9}{4}\bar{w}\frac{\mathcal{H}_{CC}}{\mathcal{H}^2\mathcal{H}_{CC}-\mathcal{H}\mathcal{H}_C^2}~~.
\end{align} 
Substituting them into the total Lagrangian, we obtain the following correction:
\begin{align}
\Delta \mathcal{L}= -\frac{81}{8}|w|^2\frac{\mathcal{H}_{CC}}{\mathcal{H}^3\mathcal{H}_{CC}-\mathcal{H}^2\mathcal{H}_C^2}=3|w|^2e^{2\mathcal{J}}\left(1-\frac{2}{3}\frac{\mathcal{J}_C^2}{\mathcal{J}_{CC}}\right).
\end{align} 
Therefore, the only effect of adding Eq.~$\eqref{constant_super}$ on the scalar potential $\eqref{potential_inflaton}$ is its modification as
\begin{align}
\nonumber V= &-\frac{e^{4\mathcal{J}/3}}{8\alpha } \left( 1- \sqrt{1+8\alpha e^{-4\mathcal{J}/3}\left( \mathcal{J}_C-\frac{\mathcal{I}}{4}e^{2\mathcal{J}/3} \right)^2} \right)\\
&-3|w|^2e^{2\mathcal{J}}\left(1-\frac{2}{3}\frac{\mathcal{J}_C^2}{\mathcal{J}_{CC}}\right).\label{potential_inflaton+w}
\end{align} 

\section{Starobinsky inflation and SUSY breaking}

In this Section we apply our model, introduced in the previous Sec.~3, to a description of cosmological inflation in supergravity, without using chiral matter supermultiplets. To be specific, we are looking for viable supergravity-based extensions of Starobinsky inflation,~\footnote{We do not provide details of Starobinsky inflation, see e.g., 
Ref.~\cite{AIKK}.} with a supersymmetry breaking vacuum after inflation.

We take the following parameterization of the functions  $\alpha$ and $\mathcal{I}$:
\begin{align}
\alpha (C)=\frac{e^{4\mathcal{J}/3M_{\rm{P}}^2}}{8M_{\rm{BI}}^4}, \ \ \mathcal{I}(C)=\xi e^{-2\mathcal{J}/3M_{\rm{P}}^2}~, \label{simplify}
\end{align} 
where $\xi$ is also $C$-dependent in general, and we have introduced the mass scale $M_{\rm{BI}}$ of the DBI structure \cite{Abe:2015fha}, in addition to the (reduced) Planck scale $M_{\rm{P}}$. Furthermore, we  restore the gauge coupling constant $g$ via the substitution $\mathcal{J}_C\rightarrow g\mathcal{J}_C$ in Eq.~$\eqref{potential_inflaton}$.

The Starobinsky-type inflation is known to be described by the following function \cite{Ferrara:2013rsa}:
\begin{align}
\mathcal{J}=-\frac{3}{2}M_{\rm{P}}^2\log \left( -\frac{C}{M_{\rm{P}}}e^{C/M_{\rm{P}}}\right)~. \label{starobinsky}
\end{align}

Substituting Eqs.~$\eqref{simplify}$ and $\eqref{starobinsky}$ into the Lagrangian $\eqref{action_inflaton}$, we find
\begin{align}
&\mathcal{L}_C =-\frac{3M^2_{\rm{P}}}{4C^2}(\partial_aC)^2 -V, \\
&V= M_{\rm{BI}}^4\left(  \sqrt{1+\frac{9g^2M^4_{\rm{P}}}{4M_{\rm{BI}}^4}\left( 1+\frac{M_{\rm{P}}}{C}+\frac{\xi}{6gM^2_{\rm{P}}}  \right)^2} -1\right)~.
\end{align}
Hence, in terms of the canonically normalized scalar $\varphi$ related to $C$ as 
$C/M_{\rm{P}}=-e^{\sqrt{\frac{2}{3}}\frac{\varphi }{M_{\rm{P}}} }$, the scalar Lagrangian is given by 
\begin{align}
&\mathcal{L}_{\varphi} =-\frac{1}{2}(\partial_a \varphi )^2 -V~,\\
&V(\varphi) = M_{\rm{BI}}^4\left(  \sqrt{1+\frac{9g^2M^4_{\rm{P}}}{4M_{\rm{BI}}^4}\left(  1-e^{-\sqrt{\frac{2}{3}}\frac{\varphi }{M_{\rm{P}}}}+\frac{\xi }{6gM_{\rm{P}}^2}\right)^2} -1\right)~.
\end{align}

We find convenient to define the dimensionless parameters as
\begin{align}
\frac{M_{\rm{P}}}{M_{\rm{BI}}}\equiv a , \ \ \frac{\xi}{M_{\rm{P}}^2}\equiv b~.
\end{align}
It is reasonable to assume that the DBI scale $M_{\rm{BI}}$ is between the GUT scale and Planck scale,
so that $a$ belongs to the interval $[1,100]$. Then the scalar potential takes the form
\begin{align}
V(\varphi) =\frac{M_{\rm{P}}^4}{a^4}\left(  \sqrt{1+\frac{9}{4}g^2a^4\left(  1-e^{-\sqrt{\frac{2}{3}}\frac{\varphi }{M_{\rm{P}}}}+\frac{1}{6g}b\right)^2} -1\right)~, \label{final_potential}
\end{align}
where the coupling constants $a$ and $b$ characterize the DBI and FI corrections, respectively. In the case of  
$g^2a^4\ll 1$ and $b/g\ll 1$ we recover the original Starobinsky model. Therefore, Eq.~(\ref{final_potential}) can be considered as the motivated two-parametric extension of Starobinsky inflationary potential in supergravity,
by using a single (massive) vector multiplet only.

If a constant superpotential is also taken into account, the corresponding scalar potential 
$\eqref{potential_inflaton+w}$  with the function (\ref{starobinsky}) reads
\begin{align}
\nonumber V(\varphi) =&\frac{M_{\rm{P}}^4}{a^4}\left(  \sqrt{1+\frac{9}{4}g^2a^4\left(  1-e^{-\sqrt{\frac{2}{3}}\frac{\varphi }{M_{\rm{P}}}}+\frac{1}{6g}b\right)^2} -1\right) \\
&-3\frac{|w|^2}{M_{\rm{P}}^2}\exp \left[ -3\sqrt{\frac{2}{3}}\frac{\varphi }{M_{\rm{P}}}+3e^{\sqrt{\frac{2}{3}}\frac{\varphi }{M_{\rm{P}}}}\right]\left( 1-\frac{2}{3}e^{\sqrt{\frac{2}{3}}\frac{\varphi }{M_{\rm{P}}}}\left(1-e^{\sqrt{\frac{2}{3}}\frac{\varphi }{M_{\rm{P}}}}\right)\right)~. \label{final_potential+w}
\end{align}
Here we observe the factor with the double exponent of the canonical scalar in the second line, indicating the dangerous "instability" of the (Starobinsky) inflation governed by the term in the first line. This phenomenon was observed in Ref.~\cite{Aldabergenov:2017hvp} in the similar context, though with a chiral (Polonyi) matter multiplet coupled to the massive vector multiplet. Because of similar "instability" (see Appendix B for details), we dispose the scalar potential (\ref{final_potential+w}) and take $w=0$ in what follows. It is worth noticing, however, that the factor with the double exponent in
(\ref{final_potential+w}) may be eliminated by changing the ${\cal J}$-function, as in Ref.~\cite{Aldabergenov:2017hvp}.

\subsection{Constant FI term}

Let us study the case of a constant coefficient at the FI term, $b=const.$ The Starobinsky-type inflationary model can be realized when $(1+\frac{1}{6g}b)> 0$.~\footnote{When $(1+\frac{1}{6g}b)< 0$, the scalar potential does not have a minimum.} The first derivative of the scalar potential $V(\varphi)$ is given by
\begin{align}
V'=\frac{9}{4}\sqrt{\frac{2}{3}}g^2M_{\rm{P}}^3e^{-\sqrt{\frac{2}{3}}\frac{\varphi }{M_{\rm{P}}}}\frac{1-e^{-\sqrt{\frac{2}{3}}\frac{\varphi }{M_{\rm{P}}}}+\frac{1}{6g}b}{\sqrt{1+\frac{9}{4}a^4g^2\left(  1-e^{-\sqrt{\frac{2}{3}}\frac{\varphi }{M_{\rm{P}}}}+\frac{1}{6g}b\right)^2}}~~,
\end{align} 
where the prime denotes the derivative with respect to $\varphi $. Demanding $V'=0$ leads to the condition
\begin{align}
1-e^{-\sqrt{\frac{2}{3}}\frac{\varphi }{M_{\rm{P}}}}+\frac{1}{6g}b=0.
\end{align}
As is clear from Eq.~$\eqref{final_potential}$, this condition results in a Minkowski vacuum at $\varphi _0/M_{\rm{P}}=-\sqrt{\frac{3}{2}}\log (1+\frac{1}{6}b)$. However, in this vacuum, we have 
\begin{align}
\left<  D\right> =\frac{3}{2}gM_{\rm{P}}^2\sqrt{\frac{\left(  1-e^{-\sqrt{\frac{2}{3}}\frac{\varphi _0}{M_{\rm{P}}} }+\frac{1}{6g}b\right)^2}{1+\frac{9}{4}a^4g^2\left(  1-e^{-\sqrt{\frac{2}{3}}\frac{\varphi _0}{M_{\rm{P}}}}+\frac{1}{6g}b\right)}}=0~~,
\end{align}
and therefore, SUSY is unbroken. This observation forces us to consider a {\it field-dependent} FI "coefficient"
$b=b(C)$ or $b=b(\varphi)$. 

\subsection{Field-dependent FI term}

When $b$ is a function of $\varphi /M_{\rm{P}}$, the critical points of the scalar potential obey the equation
\begin{align}
V'=\frac{9}{4}\sqrt{\frac{2}{3}}g^2M_{\rm{P}}^3\left( e^{-\sqrt{\frac{2}{3}}\frac{\varphi }{M_{\rm{P}}}}+\frac{M_{\rm{P}}}{2\sqrt{6}g}b' \right)\frac{1-e^{-\sqrt{\frac{2}{3}}\frac{\varphi }{M_{\rm{P}}}}+\frac{1}{6g}b}{\sqrt{1+\frac{9}{4}a^4g^2\left(  1-e^{-\sqrt{\frac{2}{3}}\frac{\varphi }{M_{\rm{P}}} }+\frac{1}{6g}b\right)^2}}=0. \label{der_V}
\end{align}
In this case, we have two equations
\begin{align}
&e^{-\sqrt{\frac{2}{3}}\frac{\varphi }{M_{\rm{P}}}}+\frac{M_{\rm{P}}}{2\sqrt{6}g}b' =0,\label{cond_1}\\
&1-e^{-\sqrt{\frac{2}{3}}\frac{\varphi }{M_{\rm{P}}} }+\frac{1}{6g}b=0.\label{cond_2}
\end{align}
Note that for Eq.~$\eqref{cond_1}$ to possess a solution, the condition $b'<0$ is required. Moreover, even when 
Eq.~$\eqref{cond_1}$ has a solution, it cannot be a true minimum because the Starobinsky potential $\eqref{final_potential}$ is non-negative, and the solution of Eq.~$\eqref{cond_2}$ always leads to a Minkowski vacuum. Hence, we consider the case when Eq.~$\eqref{cond_2}$ does {\it not\/} have solutions.

\subsubsection{Solvable case}\label{solvable_model}

As a simple example, where we can explicitly solve Eqs.~$\eqref{cond_1}$ and $\eqref{cond_2}$, let us assume that the field-dependent FI term is given by the specific function~\footnote{The same function was introduced in the similar
context in Subsec.~3.6 of \cite{Addazi:2018pbg}.}
\begin{align}
b/g=ke^{-2\sqrt{\frac{2}{3}}\frac{\varphi }{M_{\rm{P}}}}. \label{solvable_example}
\end{align}
In this case, $k>0$ is required to satisfy $(1+\frac{1}{6g}b)> 0$ and $b'<0$, which is adopted below. 

A solution to Eq.~$\eqref{cond_1}$ is given by
\begin{align}
\varphi _{*}/M_{\rm{P}} =-\sqrt{\frac{3}{2}}\log \left(  \frac{3}{k}\right) . \label{phi_star}
\end{align}
On the other hand, we have two formal solutions to Eq.~$\eqref{cond_2}$,  
\begin{align}
\varphi _{\pm}/M_{\rm{P}}=-\sqrt{\frac{3}{2}}\log \left(  \frac{3}{k}\right) -\sqrt{\frac{3}{2}}\log \left(  1\pm \sqrt{1-\frac{2}{3}k}\right) .
\end{align}
Hence, when  $\frac{3}{2}<k$, Eq.~$\eqref{cond_2}$ does not have a (real) solution. Indeed, one can show that $\varphi _{*}$ is a de Sitter minimum because of the following relations valid for $\frac{3}{2}<k$:
\begin{align}
&V|_{\varphi =\varphi _*}=\frac{M_{\rm{P}}^4}{a^4}\left( \sqrt{1+\frac{9}{4}a^4g^2\left(  1-\frac{3}{2k}\right)^2} -1\right) >0, \label{vacuum}\\
&V''|_{\varphi =\varphi _*} =\frac{9g^2M_{\rm{P}}^2}{2k}\frac{1-\frac{3}{2k}}{\sqrt{1+\frac{9}{4}a^4g^2\left(  1-\frac{3}{2k}\right)^2}} >0,\\
&\lim_{\varphi  \to \infty} V=\frac{M_{\rm{P}}^4}{a^4}\left( \sqrt{1+\frac{9}{4}a^4g^2} -1\right) , \ \ \lim_{\varphi  \to -\infty} V=\infty.
\end{align}   
At $\varphi =\varphi _*$, the vacuum expectation value of $D$ is evaluated as 
\begin{align}
\left<  D\right> =-\frac{3gM_{\rm{P}}^2}{2}\frac{1-\frac{3}{2k}}{ \sqrt{1+\frac{9}{4}a^4g^2\left(  1-\frac{3}{2k}\right)^2} }\neq 0 .
\end{align} 
Therefore, we can conclude that the minimum (vacuum) is a SUSY breaking one. As can be seen from Eq.~\eqref{vacuum}, we need $k\sim \frac{3}{2}$ to realize a tiny cosmological constant. Expanding Eq.~\eqref{vacuum} with respect to $\delta>0$, where  $k=\frac{3}{2}+\delta$, we find the following expression:
\begin{align}
V|_{\varphi =\varphi _*}=\frac{M_{\rm{P}}^4}{2}g^2\delta^2+\mathcal{O}(\delta^3).
\end{align} 
Thus we must tune our parameter $\delta$ in order to adjust the vacuum (dark) energy.

Two comments are in order. 

First, we should check the {\it no-ghost} condition, Eq.~$\eqref{cond_FF}$. During inflation, it reads 
\begin{align}
-\frac{3}{2}\left( 1-e^{-\sqrt{\frac{2}{3}}\frac{\varphi }{M_{\rm{P}}}} \right) +\left(  1+\frac{9}{4}a^4g^2\left( 1-e^{-\sqrt{\frac{2}{3}}\frac{\varphi }{M_{\rm{P}}} } \right) ^2\right)\left( 1+\frac{3}{4}ke^{-2\sqrt{\frac{2}{3}}\frac{\varphi }{M_{\rm{P}}}} \right) <0~~.
\end{align} 
Roughly speaking, the restriction $a^4g^2<\frac{2}{9}$ should be imposed, when we neglect the exponential factor 
$e^{-\sqrt{\frac{2}{3}}\frac{\varphi }{M_{\rm{P}}} }$ that is truly small during inflation.  

Second, let us check about {\it inflection} points of the scalar potential. In single-field inflationary models, an inflection point can lead to a peak in the power spectrum, that may be associated with creation of Primordial Black Holes (PBHs). In turn, the PBHs may be a (non-particle) component of dark matter \cite{Garcia-Bellido:2017mdw}.

The second derivative of our scalar potential is 
\begin{align}
V''=&-\frac{3}{2}g^2M_{\rm{P}}^2\left( 1+\frac{9}{4}a^4g^2\left(  1-x+\frac{k}{6}x^2\right)^2 \right)^{-3/2}x f(x), \label{derder_V}\\
\nonumber f(x)\equiv &\left( 1-\frac{2k}{3}x \right)\left(  1-x+\frac{k}{6}x^2\right)\left( 1+\frac{9}{4}a^4g^2\left(  1-x+\frac{k}{6}x^2\right)^2 \right)\\
&-x\left( 1-\frac{k}{3}x \right)^2  ,\label{def_fx}
\end{align}
where  
\begin{align}
x\equiv e^{-\sqrt{\frac{2}{3}}\frac{\varphi }{M_{\rm{P}}}}>0. \label{defx}
\end{align}
Hence, we are interested in solutions to $f(x)=0$. In particular, when $a,k\rightarrow 0$ (Starobinsky case), we have one such point at
\begin{align}
x=\frac{1}{2} \quad {\rm or}\quad \varphi =\sqrt{\frac{3}{2}}\log 2~. \label{infle_starobinsky}
\end{align}

For general $a$ and $k$, solving the equation $f(x)=0$ is difficult, and numerical analysis may be required. However, the latter  can be essentially avoided because we already assumed that $a$ takes its values in the interval $[1,100]$, and we derived that $k\sim\frac{3}{2}$.  As is demonstrated in Subsec.~\ref{Inflation} below, the value of $g$ is determined to be $\sim 10^{-5}$ from CMB observations. In this case, $f(x)$ becomes 
\begin{align}
f(x)=(1-x)\left( 1-x+\frac{x^2}{4}\right)\left( 1+\frac{9a^4g^2}{4}\left( 1-x+\frac{x^2}{4}\right)\right)-x\left( 1-\frac{x}{2}\right)^2, \label{nsolve}
\end{align}
where $a^4g^2$ is between $10^{-10}$ and $10^{-2}$. Then we find two solutions to $f(x)=0$ as 
\begin{align}
x=1/2,\quad 2 \quad {\rm or} \quad   \varphi =\sqrt{\frac{3}{2}}\log 2~,\quad -\sqrt{\frac{3}{2}}\log 2~, \label{sol_inflection}
\end{align}
respectively, by neglecting the term with the factor $(a^4g^2)$.~\footnote{We also solved Eq.~$\eqref{nsolve}$ numerically under the condition of $a^4g^2$ between $10^{-10}$ and $10^{-2}$, and found that there is no real solution other than Eq.~$\eqref{sol_inflection}$. } The solution $x=2$ corresponds to the vacuum, according to Eq.~$\eqref{phi_star}$. As regards another solution $x=1/2$, the first derivative of the potential,
\begin{align}
V'=\frac{9}{4}\sqrt{\frac{2}{3}}g^2M_{\rm{P}}^3\frac{x\left( 1-\frac{x}{2}\right)\left( 1-x+\frac{1}{4}x^2\right)}{\sqrt{1+\frac{9}{4}a^4g^2\left(  1-x+\frac{1}{4}x^2\right)^2}} ~~,
\end{align}
turns out to be is non-vanishing  and non-negligible at this point. Numerically, we obtain
\begin{align}
\epsilon  =0.59 \quad {\rm{for}}\quad  a^4g^2\in [10^{-10}-10^{-2}]~, 
\end{align}
at $x=1/2$, where $\epsilon$ is defined in Eq.~$\eqref{slow_para}$, while the value of $\epsilon$ is not much affected by the value of $a^4g^2$. 

Therefore, we conclude that our potential does not have an inflection point, and this excludes a formation of PBHs  in our
model.

\subsection{Cosmological parameters}\label{Inflation}

Getting an estimate of the impact of the DBI and FI corrections during inflation on the CMB observables is non-trivial. In this subsection, we briefly consider it in the particular model of Subsec.~\ref{solvable_model}.  

The relation between the number of e-foldings and inflaton filed $\varphi $ is given by
\begin{align}
N\simeq \frac{1}{M_{\rm{P}}^2}\int^{\varphi_N }_{\varphi _e}\frac{V}{V'}d\varphi \simeq \frac{3}{2}\frac{\sqrt{1+\frac{9}{4}a^4g^2}}{1+\sqrt{1+\frac{9}{4}a^4g^2}}\exp\left(\sqrt{\frac{2}{3}}\frac{\varphi_N}{M_{\rm{P}}}\right)~, \label{e-fold}
\end{align}
where $\varphi_N $ and $\varphi _e$ denote the inflaton field values at the e-foldings number $N$ and the end point of inflation, respectively. To evaluate cosmological parameters, we take the  field value of $\varphi_N$ for 
$N=50\div 60$, which is much larger than $\varphi _e$. Having obtained the leading contribution with respect to 
$\varphi_N$ on the right-hand-side of Eq.~$\eqref{e-fold}$, we can find $\varphi_N $ as a function of $N$.  

The slow-roll parameters are defined by the standard equations:
\begin{align}
\epsilon \equiv \frac{M_{\rm{P}}^2}{2}\left( \frac{V'}{V} \right)^2 \quad {\rm and} \quad \eta \equiv \frac{M_{\rm{P}}^2V''}{V}~~. \label{slow_para}
\end{align}
Using Eq.~$\eqref{e-fold}$, the values of the slow-roll parameters at $\varphi =\varphi_N $ can be rewritten as the 
functions of $N$ as follows:
\begin{align}
\epsilon_*\simeq \frac{3}{4N^2} \quad {\rm and} \quad \eta _*\simeq -\frac{1}{N}~,
\end{align}
where the subscript $(*)$ denotes the quantity evaluated at $\varphi =\varphi_N $.
Therefore, the standard CMB observables (the spectral index and the tensor-to-scalar ratio) in our case are given by
\begin{align}
&n_s=1-6\epsilon _*+2\eta _*\simeq 1 -\frac{2}{N}~,\\
&r=16\epsilon _*\simeq \frac{12}{N^2}~,
\end{align} 
in the leading order approximation. Hence, they are {\it not} affected by either of the DBI and FI parameters ($a$ and $k$). Furthermore, we have confirmed that the running of the spectral index, given by
\begin{align}
\alpha_{s*} =-2\xi_* +16\epsilon_* \eta_* -24\epsilon_* ^2\simeq -\frac{2}{N^2}, \ \ {\rm{where}}\ \ \xi \equiv  M_{\rm{P}}^4\left( \frac{V'V'''}{V^2}\right)~,
\end{align}
is {\it not} affected too, and has the same value as that in the original Starobinsky model, in the leading order approximation. 
The dependence upon $a$ and $k$, however, appears in the subleading orders, whose study is beyond the scope of this
investigation. 

The coupling constant $g$ is determined by the amplitude of the power spectrum, 
\begin{align}
A_s=\frac{V^3}{12\pi^2 M_{\rm{P}}^6V'^2} \simeq \frac{1}{18\pi^2}\frac{1}{a^4}\left( \sqrt{1+\frac{9}{4}a^4g^2}-1\right) N^2~,
\end{align}
and it is given by $A_s\sim 2\times 10^{-9}$ by CMB observations. For example, we have
\begin{align}
&(a,N)=(100,60), \ \ \Rightarrow \ \ g=9.39\times 10^{-6},\\
&(a,N)=(10,60), \ \ \Rightarrow \ \ g=9.34\times 10^{-6}.
\end{align} 

\section{Conclusion}

In this paper we studied the new supergravity model of cosmological inflation with spontaneous SUSY breaking after
inflation, beyond the standard supergravity framework, i.e. with the new FI terms that do not require gauging the R-symmetry. These FI terms significantly relax the restrictions imposed on supergravity with the standard FI term and the gauged R-symmetry and, hence, lead to the new avenues for the supergravity model building. 

By using the particular FI term, we constructed the explicit and very economical supergravity model of cosmological Starobinsky-type inflation, in terms of a single (massive) vector multiplet with the DBI structure of its kinetic terms, 
the inflaton and the Goldstino as the superpartners, and the D-type spontaneous SUSY breaking after inflation.

However, the values of the cosmological constant (the dark energy) and the SUSY breaking scale are still tightly related in
our model. It may have been expected due to the D-type of SUSY breaking used in our approach. Indeed, by using
Eqs.~(\ref{Dsol}) and (\ref{potential_inflaton}) and defining the deformation parameter $\tilde{\alpha}=8\alpha e^{-4J/3}$, we find the universal relation
\begin{align}
V= \frac{1}{\tilde{\alpha}}\left[ \frac{1}{\sqrt{1-\tilde{\alpha}D^2}}-1\right]=\frac{1}{2}D^2 + \ldots~,
\end{align}
so that  a tiny value of the cosmological constant implies a very small value of the SUSY breaking scale. This may be resolved by combining the D-type SUSY breaking  with the F-type SUSY breaking. However, this requires a separate investigation.

\section*{Acknowledgements}

HA is supported in part by a JSPS (kakenhi) Grant under No.~16K05330. YA and SVK are supported in part by the Competitiveness Enhancement Program of Tomsk Polytechnic University in Russia. SA is supported in part by a Waseda University Grant for Special Research Projects (Project number: 2018S-141). SVK is also supported in part by a Grant-in-Aid of the Japanese Society for Promotion of Science (JSPS) under No.~26400252,  and the World Premier International Research Center Initiative (WPI Initiative), MEXT, Japan. 

\begin{appendix}
\section{FI terms in curved superspace}

In this Appendix we formulate the new  FI terms in curved superspace of supergravity, by using the standard
notation and conventions of \cite{Wess:1992cp}.

\subsection{FI term I}

When employing the original (new) FI term proposed in \cite{Cribiori:2017laj}, whose coefficient $\xi$ is generalized to a function
${\cal I}(V)$, the superspace Lagrangian reads
\begin{equation}
\mathcal{L}_{\text{I}}=\mathcal{L}_{\text{mBI}}+2\int d^4\theta E\frac{W^2\bar{W}^2}{\mathcal{D}^2 W^2\bar{\mathcal{D}}^2\bar{W}^2}{\cal I}~,
\end{equation}
where the massive BI Lagrangian is given by
\begin{align}
\mathcal{L}_{\text{mBI}}=&-3\int d^4\theta Ee^{-2{\cal J}(V)/3}+\left(\frac{1}{4}\int d^2\Theta 2\mathcal{E}W^2+\text{h.c.}\right)+\nonumber\\&+\frac{1}{4}\int d^4\theta E\frac{W^2\bar{W}^2}{1+8\alpha(\omega+\bar{\omega})+\sqrt{1+8\alpha(\omega+\bar{\omega})+16\alpha^2(\omega-\bar{\omega})^2}}~~,
\end{align}
and $\omega=\frac{1}{8}\mathcal{D}^2W^2$. The ${\cal J}(C)={\cal J}(V)|$ is arbitrary real function of the real scalar $C$ that is the lowest component of the massive vector multiplet. The $\alpha$ is the BI parameter, and the vector multiplet coupling is set to one for simplicity.

After eliminating the auxiliary fields and Weyl rescaling to Einstein frame, $e\rightarrow e^{4{\cal J}/3}e$ and $g^{mn}\rightarrow e^{-2{\cal J}/3}g^{mn}$, we derive the bosonic part of the Lagrangian as follows:
\begin{align}
e^{-1}\mathcal{L}_{\text{I}}= & \frac{1}{2}R-\frac{1}{2}{\cal J}''\partial_aC\partial^aC-\frac{1}{2}{\cal J}''B_aB^a \nonumber \\
& +\frac{e^{4{\cal J}/3}}{8\alpha}\left[1-\sqrt{\displaystyle 1+8\alpha Z^2}\sqrt{\displaystyle 1+
4\alpha F^2e^{-4{\cal J}/3}+4\alpha^2(F\tilde{F})^2}\right]~,
\end{align}
where 
\begin{equation}
Z\equiv \frac{{\cal I}}{4}-{\cal J}'e^{-2{\cal J}/3}~,
\end{equation}
$\tilde{F}_{ab}\equiv-\frac{i}{2}\epsilon_{abcd}F^{cd}$, $B_a$ is the vector field whose field strength is $F_{ab}$, and the 
primes denote the derivatives with respect to $C$. The absence of ghosts requires ${\cal J}''>0$.

The auxiliary field $D$ is eliminated via its equation of motion as
\begin{equation}
D=\frac{Z}{\sqrt{1-8\alpha Z^2}}\sqrt{\displaystyle 1+4\alpha F^2e^{-4{\cal J}/3}+4\alpha^2(F\tilde{F})^2}~,
\end{equation}
and it must have the non-vanishing VEV, $\langle D\rangle\neq 0$ or $\langle Z\rangle\neq 0$, that spontaneously breaks SUSY.  The scalar potential in this case is given by
\begin{equation}
\mathcal{V}=\frac{e^{4{\cal J}/3}}{8\alpha}\left(\sqrt{1+8\alpha Z^2}-1\right)~.\label{V}
\end{equation}

\subsection{FI term II}

In the main text of our paper we employ the Lagrangian with the different FI term \cite{Kuzenko:2018jlz,Aldabergenov:2018nzd}
\begin{equation}
\mathcal{L}_{\text{II}}\supset 2\int d^4\theta E\frac{W^2\bar{W}^2}{(\mathcal{D}W)^3}{\cal I}~,
\end{equation}
where, similarly to the previous case, ${\cal I}={\cal I}(V)$. Then the D-term Lagrangian in Jordan frame reads
\begin{align}
e^{-1}\mathcal{L}_{\text{II}}(D)=&-\frac{{\cal I}}{16}\left[4D-\frac{4F^2}{D}+\frac{F^4-(F\tilde{F})^2}{D^3}\right]+e^{-2{\cal J}/3}{\cal J}'D\nonumber\\&+\frac{1}{8\alpha}\left(1-\sqrt{1+4\alpha(F^2-2D^2)+4\alpha^2(F\tilde{F})^2}\right)~.\label{LD}
\end{align}

An exact solution to the (algebraic) $D$-equation of motion amounts to finding a zero of the 5th-degree polynomial.
So, we solve it perturbatively, by ignoring the terms of $\mathcal{O}(F^4)$. Then the Lagrangian \eqref{LD} takes the form
\begin{gather}
e^{-1}\mathcal{L}_{\text{II}}(D)=-ZD+\frac{{\cal I}}{4D}F^2+\frac{1}{8\alpha}\left(1-\sqrt{1-8\alpha D^2}-\frac{2\alpha F^2}{\sqrt{1-8\alpha D^2}}\right)+{\cal O}(F^4)~.\label{LDD}
\end{gather}
We search for a solution in the form
\begin{equation}
D=D_0+D_1F^2+{\cal O}(F^4)~,
\end{equation}
and find
\begin{equation}
D_0=\frac{Z}{\sqrt{\displaystyle 1+8\alpha Z^2}}\quad {\rm and} \quad D_1=\frac{\sqrt{\displaystyle 1+8\alpha Z^2}(\frac{{\cal I}}{4}e^{-2{\cal J}/3}+2\alpha Z^3)}{Z^2+4\alpha Z^4}~.
\end{equation}
Plugging this solution into Eq.~\eqref{LDD} and Weyl-rescaling result in the full bosonic Lagrangian
\begin{equation}
e^{-1}{\cal L}_{\rm II}=\frac{1}{2}R-\frac{1}{2}{\cal J}''\partial_aC\partial^aC-\frac{1}{2}{\cal J}''B_aB^a+\frac{1}{4}\sqrt{\displaystyle 1+8\alpha Z^2}\left(\frac{{\cal I}}{Z}-1\right)F^2+{\cal O}(F^4)-{\cal V}~,\label{Lfull}
\end{equation}
where the scalar potential is
\begin{equation}
{\cal V}=\frac{\displaystyle e^{4{\cal J}/3}}{8\alpha}\left(\sqrt{\displaystyle 1+8\alpha Z^2}-1\right)~,\label{sp}
\end{equation}
i.e. {\it the same} as in the case I. This is the reason why we do not emphasize the differences between the two
FI terms in the main text of our paper, because they lead to the same scalar potentials (but the different theories).

When using Eq.~\eqref{Lfull}, we get the no-ghosts condition for $F_{ab}$ as
\begin{equation}
\frac{{\cal I}}{Z}\equiv\frac{4{\cal I}}{{\cal I}-4{\cal J}'e^{-2{\cal J}/3}}<1~.\label{ng}
\end{equation}
After the field definitions
\begin{equation}
{\cal I}=\xi e^{-2{\cal J}/3}~,~~~{\cal J}=-\frac{3}{2}\log(-Ce^C)~,~~~C=-e^{-\sqrt{2/3}\varphi}~,
\end{equation}
the condition \eqref{ng} takes the form
\begin{equation}
\frac{4\xi}{\xi-4{\cal J}'}=\frac{4\xi}{\xi+6-6e^{\sqrt{2/3}\varphi}}<1~.
\end{equation}

\section{Constant superpotential}

Let us investigate the impact of a constant superpotential in Eq.~$\eqref{final_potential+w}$ on inflation and vacuum stability in our model defined in Subsec.~\ref{solvable_model}.  The scalar potential $\eqref{final_potential+w}$ has
two parts,  
\begin{align}
V=V_{D}+V_{F}~,
\end{align}
where $V_D$ is given by Eq.~$\eqref{final_potential}$ and $V_F$ stands for the contribution of the constant superpotential. The first and second derivatives of $V_D$ are given by Eq.~$\eqref{der_V}$ subject to 
Eqs.~\eqref{solvable_example} and \eqref{derder_V}. The derivatives of $V_F$  are given by   
\begin{align}
V_F'=&-6\sqrt{\frac{2}{3}}\frac{|w|^2}{M_{\rm{P}}^2}e^{3/x}\left( 1-\frac{4}{3}x +\frac{13}{6}x^2-\frac{3}{2}x^3\right)~,\\
V_F''=&16\frac{|w|^2}{M_{\rm{P}}^2}e^{3/x}\left( -\frac{3}{4x}+1-\frac{47}{24}x+\frac{53}{24}x^2-\frac{9}{8}x^3\right)~,
\end{align}
where the field $x$ is defined by Eq.~$\eqref{defx}$.

\subsection{During inflation}

During (Starobinsky) inflation the value of $\sqrt{\frac{2}{3}}\varphi/M_{\rm P}$ varies between $5.5$ and $0.5$
\cite{AIKK}, so that $x < e^{-0.5}$.  When assuming $x \ll e^{-0.5}$, the leading contributions to $V_D$ and $V_F$ can be estimated as
\begin{align}
V_D\sim &\frac{M_{\rm{P}}^4}{a^4}\left( \sqrt{1+\frac{9}{4}a^4g^2} -1\right)~,\\
V_F\sim &-2\frac{|w|^2}{M_{\rm{P}}^2}x e^{3/x}~.
\end{align}
Hence, for the large inflaton field values the $V_F$ becomes dominant, and the derivatives of the full potential can be approximated as
\begin{align}
V'\sim -6 \sqrt{\frac{2}{3}}\frac{|w|^2}{M_{\rm{P}}^3}e^{3/x} \quad {\rm and} \quad V''\sim -12\frac{|w|^2}{M_{\rm{P}}^4}\frac{1}{x}e^{3/x}~.
\end{align}
Therefore, the slow-roll parameters, 
\begin{align}
\epsilon \sim  \frac{3}{x^2} \quad {\rm and} \quad  \eta \sim \frac{6}{x^2}~,
\end{align}
become large during the inflation, $\epsilon > 8$ and $\eta > 16$, and the slow-roll conditions are violated. 
This instability appears for any non-vanishing value of $w$.

\subsection{After inflation}

Let us examine stability of the vacuum after (Starobinsky) inflation in our model. With a non-vanishing $V_F$,
the $\varphi_*$  value deviates from that of Eq.~$\eqref{phi_star}$. We assume that a new solution to $V'=0$ takes the form $\varphi_*+\delta \varphi_*$ and then find the $\delta \varphi_*$ by solving the equation $V'=0$ in the linearized approximation. Also, for simplicity, we take $k=3/2$.  We find 
\begin{align}
V'\sim  e^{3/2} \frac{|w|^2}{M_{\rm{P}}^3}\left(10 \sqrt{6}-\frac{166}{3}\delta \varphi_* \right)+
\mathcal{O}(\delta \varphi^2_*) =0~, \quad {\rm so~that} \quad
   \delta \varphi_*\sim \sqrt{\frac{3}{2}} \frac{30}{83}M_{\rm{P}}\sim 0.4M_{\rm{P}}.
\end{align}

Inserting this solution into $V''$ yields
\begin{align}
V''|_{\varphi=\varphi_*+\delta \varphi_*}\sim g^2 M_{\rm{P}}^2 -\frac{|w|^2}{M_{\rm{P}}^4}\times \mathcal{O}(10^{2}).
\end{align}
Therefore, we find that the vacuum instability appears only for the sufficiently large $w$ when
 $|w| \geq  \frac{1}{10}  g  M^3_{\rm P}$.

\end{appendix}

\end{document}